\documentclass[draft,showkeys,showpacs,eqsecnum,nofootinbib]{revtex4}
\renewcommand{\theequation}{\arabic{equation}}
\def\beq{\begin{equation}}
\def\eeq{\end{equation}}
\def\bea{\begin{eqnarray}}
\def\eea{\end{eqnarray}}\def\nn{\nonumber}
\def\pr{\prime}
\def\prr{\prime\prime}
\def\na{\nabla}
\def\pa{\partial}

\def\nn{\nonumber}

\begin{document}
\title{Interior properties of five-dimensional Schwarzschild black hole}
\author{Soon-Tae Hong}
\email{soonhong@ewha.ac.kr}
\affiliation{Department of Science
Education and Research Institute for Basic Sciences, Ewha Womans
University, Seoul 120-750, Republic of Korea}
\date{\today}
\begin{abstract}
We investigate inner structure of Schwarzschild black hole on a five-dimensional spacetime 
$S^{3}\times {\mathbf R}^{2}$. To do this, we exploit a f\"unfbein scheme. 
In particular, we construct an equation of state of 
hydrostatic equilibrium for the five-dimensional Schwarzschild black hole, which is a 
five-dimensional version of the Tolman-Oppenheimer-Volkoff equation on four-dimensional manifold.  
We also investigate uniform density interior configuration of 
the five-dimensional black hole which consists of incompressible fluid of density, to find 
a general relativistic expression for pressure.
\end{abstract}
\pacs{04.50.Gh, 02.40.Ky, 04.20,Cv, 04.20.Jb, 04.70.Bw, 95.30.Sf}
\keywords{five-dimensional Schwarzschild black hole, f\"unfbein, interior solution, Einstein equation, 
Tolman-Oppenheimer-Volkoff equation}
\maketitle

%%%%%%%%%%%%%%%%%%%%%%%%%%%%%%%%%%%%%%%%%%%%%%%%%%%%%%%%%%%%%%%%%%%%%%%%
\section{Introduction}
\setcounter{equation}{0}
\renewcommand{\theequation}{\arabic{section}.\arabic{equation}}
%%%%%%%%%%%%%%%%%%%%%%%%%%%%%%%%%%%%%%%%%%%%%%%%%%%%%%%%%%%%%%%%%%%%%%%%

Since the five-dimensional Schwarzschild black hole metric has been proposed~\cite{tang63,emparan08}, 
tremendous progresses have been made in higher-dimensional spacetime physics. A higher-dimensional Kerr-Schild manifold possesses non-trivial parameters such as mass, angular momenta~\cite{hawking99}, while a Kerr-AdS black hole is parameterized by mass, angular momenta and cosmological constant. Here, one notes that in $D$-dimensional manifold, the number of angular momenta is equal to the rank 
of rotation group $SO(D-1)$~\cite{hashimoto04}. For positive curvature K\"ahler-Einstein manifold in 
dimension $2n$, it has been shown for a countably infinite class of associated Sasaki-Einstein manifold to exist in dimension 
$2n+3$~\cite{gaun041}. Moreover, a countably infinite number of explicit co-homogeneity one Sasaki-Einstein metrics on $S^{2}\times S^{3}$ have been presented in both the quasi-regular and irregular classes~\cite{gaun041}. It is interesting to see that, in the Kerr-AdS metric, by twisting the Killing vector field of the black holes, the compact Sasaki-Einstein manifolds constructed previously~\cite{gaun041,gaun042} have been reproduced~\cite{hashimoto04}. 

The four-dimensional Plebanski metric is characterized by a double Kerr-Schild form~\cite{page78}, where both the mass and the NUT charge are involved in the metric linearly. The most general higher-dimensional AdS-Kerr-NUT solutions have been found~\cite{hashimoto04}. This solutions can be considered as higher-dimensional generalizations of the Plebanski metric and they 
are parameterized by the mass, multiple NUT charges and arbitrary orthogonal rotations. The metric has been shown to possess $U(1)^{n}$ isometries with $n=[(D+1)/2]$. The general AdS-Kerr-NUT solutions in $D$-dimensions with $([D/2],[(D+1)/2])$ signature have been shown to admit $[D/2]$ linearly independent, mutually orthogonal and affinely-parameterized null geodesic congruences~\cite{chen07}. 

Recently, the five-dimensional Schwarzschild black hole metric has been 
exploited~\cite{tang63,emparan08},
to construct the global embedding structure, the five-acceleration and the thermodynamic physical quantity such as the Hawking temperature and entropy~\cite{hong14}. Next, the hydrodynamic properties of the five-dimensional Schwarzschild black hole have been investigated by using the massive particles and photons which are moving around the black hole, to obtain the radial component equation for the the steady state axisymmetric accretion of the massive particles and photons. The radial component of the Einstein equation associated with the entropies of the massive particles and photons has been also evaluated. On the other hand, the stringy cosmology has been investigated in higher-dimensions~\cite{hong08,hong11}. This cosmology is the string theory version of the  standard Hawking-Penrose expansion theory and it treats the twist and the shear as well as the expansion of the universe. 

In this paper, for the five-dimensional Schwarzschild black hole 
residing on the total manifold $S^{3}\times {\mathbf R}^{2}$, we investigate the five-metric in terms of the f\"unfbein, to construct the Ricci tensors and the scalar curvature. Next, we consider the equation of state related to the inner structure of the five-dimensional Schwarzschild black hole. Here we
exploit the perfect fluid stress-energy tensor to find the static, spherically symmetric solutions of Einstein equation. Finally, we investigate the Tolman-Oppenheimer-Volkoff type equation for the five-dimensional Schwarzschild black hole. This paper is organized as follows. In Section II, by 
exploiting the f\"unfbein, we set up the Ricci tensors and scalar curvature for five-dimensional Schwarzschild black 
hole metric. In Section III we investigate the interior solutions of the Einstein equation. Section IV includes the 
summaries and discussions.

%%%%%%%%%%%%%%%%%%%%%%%%%%%%%%%%%%%%%%%%%%%%%%%%%%%%%%%%
\section{Setup of Einstein equation via f\"unfbein}
\setcounter{equation}{0} 
\renewcommand{\theequation}{\arabic{section}.\arabic{equation}}
%%%%%%%%%%%%%%%%%%%%%%%%%%%%%%%%%%%%%%%%%%%%%%%%%%%%%%%%

The five-dimensional Schwarzschild black hole defined on the total
manifold $S^{3}\times {\mathbf R}^{2}$ is described in terms of the 
scalar functions $f(r)$ and $g(r)$ as follows
\beq
ds_{5}^{2}=-f(r)dt^{2}+h(r)dr^{2}+r^{2}[d\alpha^{2}+\sin^{2}\alpha
(d\theta^{2}+\sin^{2}\theta d\phi^{2})].
\label{metric5} \eeq 
Here we have three
angles of the three-sphere whose ranges are defined by $0\le
\alpha\le \pi$, $0\le \theta\le \pi$ and $0\le \phi\le 2\pi$. In the 
five-dimensional Schwarzschild black hole, the f\"unfbein is given by
\beq
\begin{array}{ll}
(e_{0})_{a}=f^{1/2}(r)(dt)_{a}, &(e_{1})_{a}=h^{1/2}(r)(dr)_{a},\\
(e_{2})_{a}=r(d\alpha)_{a}, &(e_{3})_{a}=r\sin\alpha(d\theta)_{a},\\
(e_{4})_{a}=r\sin\alpha\sin\theta(d\phi)_{a}, &\label{funf}
\end{array}
\eeq
and the inverse f\"unfbein is also described as follows
\beq
\begin{array}{ll}
(e_{0})^{a}=-f^{-1/2}(r)(\pa_{t})^{a}, &(e_{1})^{a}=h^{-1/2}(r)(\pa_{r})^{a},\\
(e_{2})^{a}=r^{-1}(\pa_{\alpha})^{a}, &(e_{3})_{a}=(r\sin\alpha)^{-1}(\pa_{\theta})^{a},\\
(e_{4})^{a}=(r\sin\alpha\sin\theta)^{-1}(\pa_{\phi})^{a}. &\label{funfinv}
\end{array}
\eeq
The f\"unfbein structure will be also exploited in the next section to investigate the equation of state of 
hydrostatic equilibrium for the five-dimensional Schwarzschild black hole, namely the five-dimensional 
Tolman-Oppenheimer-Volkoff type equation. Moreover, we have the identities among the f\"unfbein and inverse f\"unfbein
\beq
(e_{\mu})^{a}~(e_{\nu})_{a}=\eta_{\mu\nu},~~~(e_{\mu})^{a}~(e_{\nu})^{b}~\eta^{\mu\nu}=g^{ab},~~~
(e_{\mu})^{a}~(e_{\nu})_{b}~\eta^{\mu\nu}=\delta^{a}_{b}.
\eeq
Exploiting the above metric (\ref{metric5}), after some algebra we obtain the 
Ricci tensors $R_{ab}$ $(a,b=0,1,2,3,4)$ as follows
\bea
R_{00}&=&\frac{f^{\prr}}{2h}-\frac{(f^{\pr})^{2}}{4fh}-\frac{f^{\pr}h^{\pr}}{4h^{2}}+\frac{3f^{\pr}}{2rh},\nn\\
R_{11}&=&-\frac{f^{\prr}}{2f}+\frac{(f^{\pr})^{2}}{4f^{2}}+\frac{f^{\pr}h^{\pr}}{4fh}+\frac{3h^{\pr}}{2rh},\nn\\
R_{22}&=&-\frac{rf^{\pr}}{2fh}+\frac{rh^{\pr}}{2h^{2}}+2-\frac{2}{h},\nn\\
R_{33}&=&\left(-\frac{rf^{\pr}}{2fh}+\frac{rh^{\pr}}{2h^{2}}+2-\frac{2}{h}\right)\sin^{2}\alpha,\nn\\
R_{44}&=&\left(-\frac{rf^{\pr}}{2fh}+\frac{rh^{\pr}}{2h^{2}}+2-\frac{2}{h}\right)\sin^{2}\alpha\sin^{2}\theta.
\label{ricci}
\eea 
Here the primes denote the time derivatives. With the Ricci tensors in mind, we find the Einstein equation
\beq
G_{ab}=R_{ab}-\frac{1}{2}R=8\pi T_{ab},
\label{einstein}
\eeq 
where the scalar curvatute $R$ is given by 
\beq
R=-\frac{f^{\prr}}{fh}+\frac{(f^{\pr})^{2}}{2f^{2}h}+\frac{f^{\pr}h^{\pr}}{2fh^{2}}-\frac{3f^{\pr}}{rfh}
+\frac{3h^{\pr}}{rh^{2}}+\frac{6}{r^{2}}-\frac{6}{r^{2}h}
\eeq
and $T_{ab}$ is the fluid stress-energy tensor which will be described in the next section in detail. 

For the sake of completeness and ensuing uses in the next section, we first consider the vacuum solutions of 
the Einstein equation with $T_{ab}=0$,
\beq
R_{ab}=0,
\eeq 
outside the five-dimensional Schwarzschild black hole. By adding the 
Ricci tensors $(1/f)R_{00}$ and $(1/h)R_{11}$ we obtain
\beq
\frac{df}{f}+\frac{dh}{h}=0,
\eeq
to yield with a constant $C_{1}$
\beq
f=\frac{C_{1}}{h}=\frac{1}{h}. 
\label{fsol1}
\label{fkh}
\eeq
In the second equality in (\ref{fkh}), we have used the fact that by rescaling the time coordinate, 
$t\rightarrow C_{1}^{1/2}t$, we may set $C_{1}=1$. Next, $R_{22}=0$ in (\ref{ricci}) produces
\beq
\frac{d}{dr}(r^{2}f)-2r=0,
\eeq
which yields
\beq
f=1+\frac{C_{2}}{r^{2}}.
\label{fsol2}
\eeq
We next consider the behavior of a test body in the five-dimensional Newtonian gravitational 
field of mass $M$~\cite{emparan08}, to fix the coefficient $C_{2}$ as follows: $C_{2}=-\beta M$
with a new constant $\beta$
\beq
\beta=\frac{8}{3\pi}.
\label{beta}
\eeq
Substituting $f$ and $h$ in (\ref{fsol1}), (\ref{fsol2}) and (\ref{beta}) into (\ref{metric5}), 
we obtain the vacuum solution outside the five-dimensional Schwarzschild black hole, whose metric is of the form
\beq
ds_{5}^{2}=-\left(1-\frac{\beta M}{r^{2}}\right)dt^{2}+\left(1-\frac{\beta M}{r^{2}}\right)^{-1}dr^{2}
+r^{2}[d\alpha^{2}+\sin^{2}\alpha(d\theta^{2}+\sin^{2}\theta d\phi^{2})].
\label{metric52} \eeq 
Here one notes that at $r=0$ we have a physical singularity while at $r=r_{S}$ we have a coordinate singularity where
$r_{S}$ is given by
\beq
r_{S}=(\beta M)^{1/2}.
\eeq

%%%%%%%%%%%%%%%%%%%%%%%%%%%%%%%%%%%%%%%%%%%%%%%%%%%%%%%%
\section{Interior solutions of Einstein equation}
\setcounter{equation}{0}
\renewcommand{\theequation}{\arabic{section}.\arabic{equation}}
%%%%%%%%%%%%%%%%%%%%%%%%%%%%%%%%%%%%%%%%%%%%%%%%%%%%%%%%

In this section, we will investigate the equation of state associated with the 
inner structure of the Schwarzschild black hole in the five-dimensional total 
manifold. We observe here that the details of the 
ingredients such as the Ricci tensors and Einstein tensors in the five-dimensional spacetime 
are quite nontrivially different from those in the four-dimensional one, as shown in (\ref{ricci}) and 
(\ref{etensor1})-(\ref{etensor3}) below.  

We now find the static and spherically symmetric solutions of Einstein equation. 
To do this, we assume the perfect fluid stress-energy tensor of the form
\beq
T_{ab}=\rho u_{a}u_{b}+P(g_{ab}+u_{a}u_{b}),
\label{tab}
\eeq
where $\rho$ and $P$ are the mass-energy density and pressure of the fluid.
Here one notes that the fluid four-velocity $u^{a}$ should point in the same 
direction as the static Killing vector field, to yield
\beq
u^{a}=-(e_{0})^{a}=f^{-1/2}(\pa_{t})^{a}
\eeq
The perfect fluid stress-energy tensor in (\ref{tab}) with flat indices $(A,B=0,1,2,3,4)$ then becomes
\beq
T_{AB}={\rm diag}~(\rho, P, P, P, P).
\label{tabflat}
\eeq
Substituting (\ref{tabflat}) into (\ref{einstein}), we next obtain the three independent equations
\bea
8\pi\rho&=&G_{00}f^{-1}=\frac{3h^{\pr}}{2rh^{2}}+\frac{3}{r^{2}}-\frac{3}{r^{2}h},\label{etensor1}\\
8\pi P&=&G_{11}h^{-1}=\frac{3f^{\pr}}{2rfh}-\frac{3}{r^{2}}+\frac{3}{r^{2}h},\label{etensor2}\\
8\pi P&=&G_{22}r^{-2}=\frac{f^{\pr}}{rfh}-\frac{h^{\pr}}{rh^{2}}+\frac{f^{\prr}}{2fh}
-\frac{(f^{\pr})^{2}}{4f^{2}h}-\frac{f^{\pr}h^{\pr}}{4fh^{2}}-\frac{1}{r^{2}}+\frac{1}{r^{2}h}.
\label{etensor3}
\eea
Here one notes that $G_{33}$ and $G_{44}$ reproduce the result in (\ref{etensor3}), since we have 
\beq
G_{33}=G_{22}\sin^{2}\alpha,~~~G_{33}=G_{22}\sin^{2}\alpha\sin^{2}\theta.
\eeq

Noticing that the equation in (\ref{etensor1}) contains only $h$, we shuffle 
this $\rho$-equation as follows
\beq
8\pi\rho=\frac{3}{2r^{3}}\frac{d}{dr}\left[r^{2}\left(1-\frac{1}{h}\right)\right],
\eeq
which yields
\beq
h(r)=\left[1-\frac{\beta m(r)}{r^{2}}\right]^{-1}.
\label{hr}
\eeq
Here $\beta$ is the constant in (\ref{beta}) and $m(r)$ is a mass in the five-dimensional spacetime defined by
\beq
m(r)=2\pi^{2}\int_{0}^{r}\rho(r^{\pr})r^{\pr~3}dr^{\pr}.
\label{mrint}
\eeq
Moreover, we observe that a necessary condition for staticity is $h\ge 0$ to yield $r\ge [\beta m(r)]^{1/2}$.
Assuming that $\rho=0$ for $r>R$, we notes that the interior solution $h$ in (\ref{hr}) joins on the vacuum solution 
(\ref{metric52}) with the total mass $M$ as follows
\beq
M=m(R)=2\pi^{2}\int_{0}^{R}\rho(r)r^{3}dr,
\eeq
which is identical to the form for the total mass in Newtonian gravity in the five-dimensional spacetime. 
The total proper mass $M_{p}$ should be now defined on the proper volume element
\beq
[^{(4)}g]^{1/2}d^{4}x=h^{1/2}r^{3}\sin^{2}\alpha\sin\theta~dr~d\alpha~d\theta~d\phi,
\eeq
to yield 
\beq
M_{p}=2\pi^{2}\int_{0}^{R}\rho(r)\left[1-\frac{\beta m(r)}{r^{2}}\right]^{-1/2}r^{3}dr
\eeq
The difference $E_{B}=M_{p}-M$ can be interpreted as the gravitational binding energy of the configuration.

Now we assume that $f$ is given by the following form
\beq
f=e^{\beta\phi}
\eeq
then, from (\ref{etensor2}), we obtain
\beq
\frac{d\phi}{dr}=\frac{2[m(r)+\pi^{2}r^{4}P]}{r[r^{2}-\beta m(r)]}.
\label{dphidr}
\eeq
In the Newtonian limit where $r^{4}P\ll m(r)$ and $m(r)\ll r^{2}$, (\ref{dphidr}) reduces to
\beq
\frac{d\phi}{dr}\cong \frac{2m(r)}{r^{3}}.
\eeq
Next, we need to find the equation of state inside the given black hole by exploiting the remnant equation (\ref{etensor3}). However 
it is not so trivial to manipulate (\ref{etensor3}) in order to arrive at the desired equation of state. Instead, we use the 
connection one-form $\omega_{a\mu\nu}$ defined in terms of the f\"unfbein and 
inverse f\"unfbein in (\ref{funf}) and (\ref{funfinv}), respectively, as follows
\beq
\omega_{a\mu\nu}=(e_{\mu})^{b}\na_{a}(e_{\nu})_{b}=(e_{\mu})^{b}\omega_{ab\nu}.
\eeq
Exploiting the Einstein equation in (\ref{einstein}), we readily obtain 
\beq
(e_{\mu})_{b}~8\pi\na_{a}T^{ab}=\omega_{ab\mu}(G_{ab}-8\pi T_{ab})-\na_{a}[(e_{\mu})_{b}(G^{ab}-8\pi T^{ab})],
\eeq
which implies that the Einstein equation (\ref{einstein}) is equal to 
\beq
\na_{a}T^{ab}=0.
\label{natab}
\eeq
The identity in (\ref{natab}) produces the following two equations
\bea
(\rho+P)u^{a}\na_{a}u_{b}+(g_{ab}+u_{a}u_{b})\na^{a}P=0,\label{id1}\\
u^{a}\na_{a}\rho+(\rho+P)\na^{a}u_{a}=0,\label{id2}
\eea
which hold regardless of the dimensionality of the given target manifold. Exploiting (\ref{id1}), we arrive at an 
equation of state of hydrostatic equilibrium for the five-dimensional Schwarzschild black hole as follows
\beq
\frac{dP}{dr}=-\beta(\rho+P)\frac{m(r)+\pi^{2}r^{4}P}{r[r^{2}-\beta m(r)]},
\label{dpdr}
\eeq
where we have used a Killing vector condition
\beq
\na_{a}u_{b}+\na_{b}u_{a}=0,
\eeq
and the following relations
\bea
(e_{1})_{b}~u^{a}\na_{a}u_{b}&=&\frac{1}{2}\beta h^{-1/2}\phi^{\pr},\nn\\
(e_{1})_{b}~g^{ab}\na_{a}P&=&h^{-1/2}\pa_{r}P.
\eea
Here we note that, in the four-dimensional black hole, the corresponding equation of state is called 
Tolman-Oppenheimer-Volkoff equation and is given by~\cite{tolman39,oppenheimer39,wald}
\beq
\frac{dP}{dr}=-(\rho+P)\frac{m(r)+4\pi r^{3}P}{r[r^{2}-2m(r)]}
\eeq 
In the Newtonian limit where $r^{4}P\ll m(r)$ and $m(r)\ll r^{2}$, (\ref{dpdr}) reduces to
\beq
\frac{dP}{dr}\cong -\beta\rho\frac{m(r)}{r^{3}}.
\label{dpdr2}
\eeq
Here one notes that the spacetime metric inside the five-dimensional Schwarzschild black hole is given by
\beq
ds^{2}=-e^{\beta\phi}dt^{2}+\left[1-\frac{\beta m(r)}{r^{2}}\right]^{-1}dr^{2}+r^{2}[d\alpha^{2}+\sin^{2}\alpha
(d\theta^{2}+\sin^{2}\theta d\phi^{2})].
\eeq

%%%%%%%%%%%%%%%%%%%%%%%%%%%%%%%%%%%%%%%%%%%%%%%%%%%%%%%%%%%%%%%%%%%%%%%%
\section{Conclusions and discussions}
\setcounter{equation}{0}
\renewcommand{\theequation}{\arabic{section}.\arabic{equation}}
%%%%%%%%%%%%%%%%%%%%%%%%%%%%%%%%%%%%%%%%%%%%%%%%%%%%%%%%%%%%%%%%%%%%%%%%

In summary, we have investigated the inner structure of the Schwarzschild black hole defined  
on the five-dimensional spacetime. More specifically, we have obtained the equation of 
state of hydrostatic equilibrium for the five-dimensional Schwarzschild black hole by exploiting 
the f\"unfbein scheme. We have formulated the Einstein equation for this five-dimensional 
Schwarzschild black hole. To do this, we have found the Ricci tensors and scalar curvature 
for the five-dimensional target manifold. Using the Einstein equation, we have constructed 
the interior solutions for the five-dimensional Schwarzschild black hole and the corresponding 
the spacetime metric inside the black hole.

We have several comments to address. First, we consider uniform density interior configuration of 
the five-dimensional black hole which consists of incompressible fluid of density $\rho_{0}$:
\beq
\rho(r)=\left\{\begin{array}{ll}
\rho_{0} &(r\le R) \\
0 &(r>R),
\end{array}\right.
\label{rhor}
\eeq
from which exploiting (\ref{mrint}) we obtain 
\beq
m(r)=\frac{1}{2}\pi^{2}\rho_{0}r^{4}.
\label{mrrho}
\eeq
In the Newtonian limit where $r^{4}P\ll m(r)$ and $m(r)\ll r^{2}$, (\ref{dpdr2}) yields  
\beq
P(r)\cong \frac{1}{4}\beta\pi^{2}\rho_{0}^{2}(R^{2}-r^{2}),
\eeq
which is consistent with the boundary condition $P(R)\cong 0$. Moreover the pressure at the core is readily obtained to yield
\beq
P_{c}\cong \frac{1}{4}\beta\pi^{2}\rho_{0}^{2}R^{2},
\label{pcnew}
\eeq
which is also rewritten as follows
\beq
P_{c}\cong\frac{1}{2^{3/2}}\beta\pi\rho_{0}^{3/2}M^{1/2}.
\label{pcnew2}
\eeq
Here we have used the identity
\beq
R=\left(\frac{2M}{\pi^{2}\rho_{0}}\right)^{1/4}.
\eeq
After some algebra together with (\ref{mrrho}), the general relativistic expression for $P$ is then given by 
\beq
P(r)=\rho_{0}\frac{(1-\beta M/R^{2})^{1/2}-(1-\beta M r^{2}/R^{4})^{1/2}}{(1-\beta M r^{2}/R^{4})^{1/2}-2(1-\beta M/R^{2})^{1/2}},
\eeq
which yields the pressure at the core of the black hole defined in the four-dimensional spacetime 
\beq
P_{c}=\rho_{0}\frac{(1-\beta M/R^{2})^{1/2}-1}{1-2(1-\beta M/R^{2})^{1/2}}.
\label{pc3}
\eeq
Here one can readily observe that, in the Newtonian limit $P_{c}$, the central pressure 
in (\ref{pc3}) reduces to (\ref{pcnew}).

\end{document}